# Measurement of the molecular dipole moment and the hyperfine and Λ-doublet splittings of the $B^3\Pi_1$ state of thallium fluoride

N. B. Clayburn[1], T. H. Wright[1,2,*], E. B. Norrgard[3], D. DeMille[2,†], L. R. Hunter[1]

[1]*Physics Department, Amherst College, Amherst, Massachusetts 01002, USA*
[2]*Department of Physics, P.O. Box 208120, Yale University, New Haven, Connecticut 06520, USA*
[3]*Joint Quantum Institute, National Institute of Standards and Technology and University of Maryland, Gaithersburg, Maryland 20899, USA*

## ABSTRACT

We report high-precision measurements on the thallium fluoride $\tilde{J} = 1$ hyperfine manifold of the $B^3\Pi_1$ ($v = 0$) state. This state is of special interest because it is central to an optical cycling scheme that is envisioned to play an important role in enhancing the sensitivity of the CeNTREX nuclear Schiff-moment experiment presently under construction. The measurements are made by monitoring the fluorescence induced by narrow-band laser excitation of a cryogenic molecular beam. We use a multipass arrangement of the laser beam to enhance fluorescence. When viewed with a camera, we can spatially resolve images from adjacent passes that approach the molecules from opposing directions. These images yield a sensitive visual method to identify the central frequency of a transition. Coupling these line-center determinations with frequency calibration from an acousto-optic modulator has allowed a more precise determination of the $\tilde{J} = 1$ manifold of hyperfine level splittings. We observe Stark shifts of the $\tilde{J} = 1$ levels and infer a permanent electric dipole moment of 2.28(7) D and Λ-doublet splittings for the $F_1' = 1/2$ and $F_1' = 3/2$ manifolds of 14.4(9) and 17.4(11) MHz, respectively.

## I. INTRODUCTION

The thallium fluoride (TlF) $X^1\Sigma^+$ state has been previously used to make precision tests of parity- and time-reversal symmetry violations [1-3]. The high mass of Tl and the high polarizability of the molecule make TlF ideal for measuring the Schiff moment of the Tl nucleus [4]. The TlF $B^3\Pi_1$ ($v_e = 0$) ← $X^1\Sigma^+$ ($v_g = 0$) transition has been proposed [5] as a candidate for optical cycling and laser cooling; such techniques could be effective for enhancing the sensitivity of symmetry violation measurements [6-8]. Here, $v_g$ and $v_e$ are the ground- and excited-state vibrational quantum number. Laser cooling and cycling as a means to enhance symmetry violation measurements has been proposed in other diatomic molecules such as BaF [9], RaF [10], and YbF [11,12] as well as polyatomic molecules like BaOH and YbOH [13,14].

[*]Present address: JILA, National Institute of Standards and Technology and University of Colorado, Boulder, Colorado 80309, USA; Department of Physics, University of Colorado, Boulder, Colorado 80309, USA.
[†]Present address: James Franck Institute and Department of Physics, The University of Chicago, Chicago, Illinois 60637, USA; Physics Division, Argonne National Laboratory, Argonne, Illinois 60439, USA.

The TlF $B^3\Pi_1$ state has resolved hyperfine (HF) structure. The HF interaction produces mixing of states with different rotational quantum numbers, $J$. This mixing can spoil the usual rotational selection rules and lead to branching to additional ground rotational levels, thus compromising optical cycling. Therefore, in order to achieve optical cycling, it is critical to understand the rotational and HF structure of the excited states and their effects on rotational branching.

In earlier work, the $X$-state HF and rotational energies were determined by high-resolution microwave spectroscopy [15] and the rovibrational energies of the $B^3\Pi_1$ state were determined by low-resolution spectroscopy with a pulsed UV laser [16]. Recently, high-resolution laser spectroscopy of the $B^3\Pi_1$ ($v_e = 0$) ← $X^1\Sigma^+$ ($v_g = 0$) transition resolved the excited-state HF structure and inferred the parameters describing the HF interaction of the $B$ state [17,18]. The Hamiltonian parameters that describe the $B$ state are derived from the data of Ref. [17] using the effective Hamiltonian of Ref. [18]. The prior analysis of Ref. [17] incorrectly accounted for lambda-doubling effects, so the Hamiltonian and parameters of Ref. [18] should be taken to supersede the prior result. In this work we describe experiments which measure the permanent electric dipole moment (EDM) and the Λ-doublet splitting of the $B^3\Pi_1$ state. We also describe a technique for measuring HF splittings that has yielded more precise values for some of the critical hyperfine intervals.

## II. APPARATUS

The apparatus consists of a cryogenic buffer gas beam source which produces a collimated molecular beam. This molecular beam is crossed with resonant laser light and fluorescence is observed perpendicular to these intersecting beams. We detect this fluorescence with either a camera or a phototube.

### A. Vacuum apparatus

In detail, we use a cryogenic buffer gas beam source similar to that of Ref. [19] and described further in Ref. [17]. A solid target of TlF is made by melting TlF powder in a copper crucible. The filled crucible is fixed to a copper cell which is mounted to a two-stage pulse tube refrigerator (CryoMech, PT415) [20]. The target is held at ~4 K and TlF molecules are produced by laser ablation of the target by intense 1064-nm light from a Nd:YAG laser (Big Sky Laser Technologies, Ultra GRM). The Nd:YAG (yttrium aluminum garnet) laser produces 10-ns, 30-mJ pulses with a 1.4-Hz repetition rate that is synchronized with the pulse tube cooling cycle. A flow of cryogenic helium buffer gas thermalizes the ablated molecules and then extracts the molecules from the cell, directing the molecular beam though a 6.35 mm hole to a ~$10^{-7}$ Torr science chamber where experiments are performed.

Inside the science chamber is a 1.98 mm by 6.35 mm horizontally oriented collimating slit. This collimating slit is 30 cm from the molecular source and is 6.35 cm away from the center of the interaction region. Centered above and below this interaction region are a pair of 7.62-cm-diameter circular parallel-plate electrodes (Fig. 1). The polished electrodes are separated by 3.175 cm and are used to produce a uniform electric field, ranging from 0-300 V/cm. Each electrode is 6.35 mm thick. Five hundred and eighty-three 1.32-mm-diameter holes are machined into the upper electrode plate to allow detection of laser-induced fluorescence from the interaction region. This hole pattern (70% normal incidence transmission) spans a centered circular region with a diameter of 3.81 cm. This circular region is recessed such that the holes have a depth of 1.52 mm. This recess, which faces away from the interaction region, serves to reduce the depth of the holes for easier fabrication and does not significantly change the homogeneity of the field or the distance between electrodes. Both the collimating plate and the electrodes are made of brass and coated with 15±3 μm of gold. This prevents oxidization which could leave nonconducting patches where stray charge might accumulate and compromise the electric-field uniformity. The nearest equipotential surface to the interaction region, besides the electrodes, is the fixture holding the collimating slit which is 5.5 cm away.

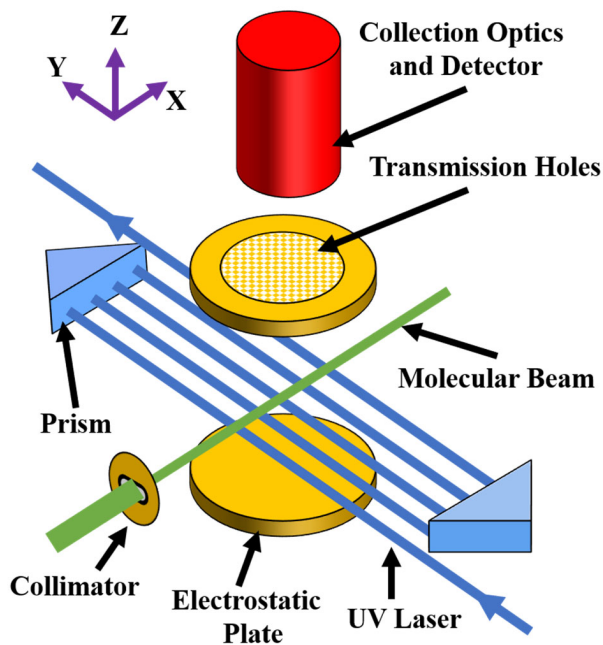

**FIG. 1**. Schematic of the interaction region. Each pass of the laser (±y direction) is perpendicular to the propagation direction of the molecular beam (+x direction). Laser-induced fluorescence is observed in the same direction as the imposed electric field (+z direction). The figure is not to scale.

### B. Excitation light

The $B(v_e = 0) \leftarrow X(v_g = 0)$ transition occurs at 271.7 nm. A narrow, tunable, 1087-nm fiber laser (Keopsys, CYFL-KILO) is frequency doubled twice using commercial Toptica bow-tie cavities. This system produces ~20 mW of 271.7-nm cw light. The infrared fiber-laser frequency is locked to a scanning Fabry-Pérot cavity by monitoring its transmission. The length of the cavity is maintained by simultaneously monitoring the transmission of a frequency-stabilized helium-neon laser (Laboratory for Science, model 210).

The laser light intersects the collimated molecular beam perpendicular to the direction of molecular motion (Fig. [1]). Specifically, two antireflection-coated quartz right-angle prisms are aligned so that the laser light bounces through the interaction region between 5 and 11 times [21]. This "multipass configuration" is used to maximize the time over which each TlF molecule interacts with the light. The laser has a $1/e^2$ radius of 0.6 mm and the multipass extends 18 mm along the molecular-beam path. The width of the molecular beam at the interaction region is about 7 mm. Careful alignment of the multipass results in beams which are all parallel (or antiparallel) to one another and perpendicular to the molecular beam, thus minimizing the Doppler broadening of the molecular transition.

### C. Fluorescence detection and techniques

Molecules excited by the incident laser light decay back to the ground state with a characteristic lifetime of 99(9) ns [5]. Some of the resulting laser-induced fluorescence passes through holes in the upper electrode plate, through a vacuum window, and into imaging optics. The imaging optics consists of a quartz collection and collimating lens, a narrow-band interference filter, and a quartz focusing lens which directs the light to a detector. The fluorescence is detected by either a photomultiplier tube, (Hamamatsu, R928) or a UV-sensitive camera (Princeton Instruments, PiMax 2). The PMT has superior signal to noise, while the camera allows for spatial resolution and imaging of the individual multipass beams.

The ablated TlF molecules reach the multipass region ~10 ms after the Nd:YAG fires. When using the PMT, fluorescence signals are recorded for 50 ms immediately after this firing. We report in this work the integrated fluorescence signals collected for 20 ms, beginning 10 ms after the firing of the Nd:YAG, this integrated signal containing most of the fluorescence signal. Typically, we record this PMT fluorescence signal as a function of the laser frequency to trace out spectroscopic peaks of the relevant transitions. When using the UV-sensitive camera, we image the interaction region over a similar 20-ms exposure time, and subtract an image with equal exposure in the absence of molecules. We combine multiple background-subtracted images to reach the desired signal size for a given experimental configuration. These images have been empirically corrected to account for the spatially varying detection efficiency of our collection optics. A biquadratic function maximized at the image center is used to describe the monotonic drop in detection efficiency along the long dimension ($x$ direction) of the images. The detection efficiency varies by <40% over the region where the molecular fluorescence is seen, with the majority of this variation occurring near the edges of the images. The imaging techniques of this work rely primarily on comparing the spatial location of the fluorescence from various laser passes. As such, these detection efficiency-driven corrections to the fluorescence intensity have only a peripheral effect on the conclusions drawn from the images.

While the natural decay rate of the $B$ state is $\Gamma = 1\times10^7$ s$^{-1}$, polarization and hyperfine dark states of the $X(J = 1)$ ground state dramatically reduce the photon cycling rates compared with those of a two-level system [6]. Because the ground-state hyperfine structure of TlF is unresolved, when one excites to a single fully resolved upper-state hyperfine level, the exciting laser couples to at most a single coherent superposition of the ground-state hyperfine manifold for each total angular momentum projection $m_F$. We refer to this superposition as the "hyperfine bright state," while the other orthogonal linear superpositions are the "hyperfine dark states." Typically, these hyperfine dark states evolve into bright states at a rate determined by the ground-state hyperfine splitting. This rate is of order ~$2\pi\times10$ kHz to $2\pi\times100$ kHz for TlF, much smaller than the natural decay rate of the $B$ state.

We increase the rate of cycling out of these dark states by rapidly switching the exciting laser's polarization with an electro-optical modulator [22] and by resonantly driving the microwave transition between the $J = 0$ and $J = 1$ rotational ground states [6]. The laser and microwave polarizations are switched rapidly (typically ~1 MHz) and are modulated 90° out of phase with one another. For transitions which are highly closed (to unwanted electronic, vibrational, and rotational decay paths), using these techniques drastically increases the number of photons scattered per molecule and increase our signal-to-noise ratio. Even in the absence of a closed transition, population transfer from the $X(J = 0)$ ground state due to resonantly tuned microwaves can increase fluorescence and similarly improve signal to noise. The Stark shifts associated with the microwaves have been theoretically estimated and they are found to be negligible at our level of accuracy. Empirically, the microwaves do not induce an observable shift in the fluorescence line centers.

### III. THEORY

#### State notation

Fluorine has only one isotope, $^{19}$F, and although thallium has two commonly occurring isotopes, $^{203}$Tl and $^{205}$Tl, this work investigates only the latter. The theoretical description of the $B$ state of TlF is detailed in Refs. [17] and [18]. We describe it briefly here. The state is described by the Hund's case (c) basis and the coupling scheme:

$$\boldsymbol{F}_1 = \boldsymbol{J} + \boldsymbol{I}_1, \quad (1)$$
$$\boldsymbol{F} = \boldsymbol{F}_1 + \boldsymbol{I}_1. \quad (2)$$

Here the thallium nuclear spin is $I_1 = 1/2$, the fluorine nuclear spin is $I_2 = 1/2$, and the total angular momentum of the molecules less nuclear spin is $J$. The HF states associated with quantum number $J$ are then $F_1 = J \pm 1/2$ and $F = J - 1, J,$ and $J + 1$. The basis kets for states of parity $P = \pm 1$ are

$$|c\rangle = |J, \Omega, I_1, F_1, I_2, F, m_F, P\rangle, \quad (3)$$

where $\Omega$ is the projection of $J$ on the internuclear axis and $m_F$ is the projection of $F$ along $z$ in the laboratory frame. Following the convention of Herzberg [23], we refer to states with $P = (-1)^J$ as $e$-parity and $P = (-1)^{J+1}$ as $f$-parity states. Here the rotational quantum number $J$ describes states in the above basis. In order to account for HF interactions mixing neighboring rotational levels in the excited $B$ state, we label $B$-state energy eigenstates with the approximate quantum number $\tilde{J}$. We denote ground $X$-state energy eigenstates with $J_g$. Excited-state quantum numbers are primed to distinguish them from ground-state quantum numbers.

### IV. METHODS AND RESULTS

#### A. AOM coincidence measurements

Spectroscopy of the $B$ state from Ref. [17] determined many line positions by referencing the tunable laser to a frequency-stabilized HeNe laser via a scanning Fabry-Pérot cavity transfer lock. There, the largest HF splittings in the $\tilde{J} = 1$ manifold were limited to an accuracy of ~20 MHz, primarily due to nonlinearity in the cavity scan. We use the same system in this work. Furthermore, we have found that the roughly 10-MHz long-term drift in the reference HeNe laser limits our ability to compare line positions over timescales greater than about an hour. Here, we use the technique of acousto-optic modulator (AOM) coincident resonance outlined below to determine line splittings more accurately. This method references the measured line splitting to a high-stability rf source used to drive an AOM, avoiding the predominant systematic uncertainties associated with the cavity transfer lock. Additionally, we use spatial fluorescence information from our camera to tune the laser to resonance with the zero transverse velocity class of the molecular beam. With this technique, we measured five $\tilde{J} = 1$ splittings with 2-MHz precision. The electrodes were removed from the vacuum chamber for the AOM coincident resonance measurements in order to provide unobstructed imaging with the camera.

The AOM coincident resonance technique relies on the geometric properties of the laser multipass, which cause each subsequent laser pass to propagate in the opposite direction. Ultimately, this results in counterpropagating neighboring passes as shown in Fig. 2. This, coupled with the molecular beam's transverse expansion, results in different molecular velocity classes fluorescing at different points along the multipass. Specifically, when the laser's frequency is below (above) resonance, the molecules traveling toward (away from) the laser beam are on resonance and get preferentially excited. Because our camera and collection optics allow us to spatially resolve each multipass beam, we can see this effect and use it as a pictorial resonance condition. This yields substantially better resolution than is possible from our Doppler-broadened laser-induced fluorescence spectra. Background-subtracted images averaged over multiple shots from this process are shown in Fig. 3.

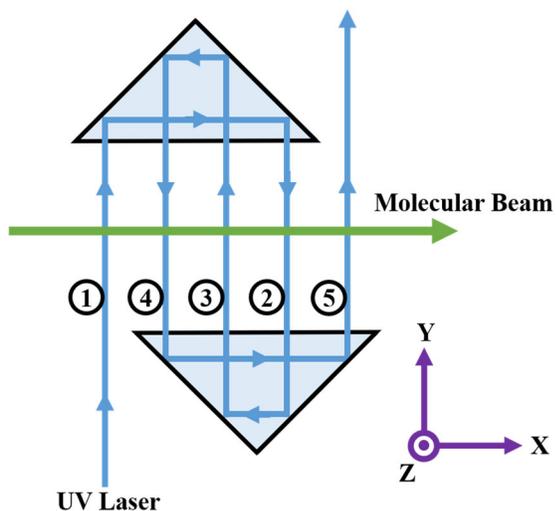

**FIG. 2.** Schematic of the multipass viewed from above with the passes numbered. From this vantage point, consider laser-induced molecular fluorescence by a red detuned laser. For odd-numbered passes, in which the laser beam propagates in the $+y$ direction, molecules with a $-y$-direction velocity component will preferentially fluoresce. Such molecules are concentrated near the bottom due to the molecular beam's transverse expansion. For even-numbered passes, traveling in the $-y$ direction, molecules near the top will preferentially fluoresce [(see Fig. 3c)].

Consider, for example, the determination of the splitting between the $Q_1$, $F_1' = 1/2$, $F' = 0,1$ transitions using this AOM-based technique. When a laser passes through an AOM under the Bragg condition two beams are output, a zeroth-order beam at the same frequency as the input beam, and a first-order beam which has been shifted in frequency. This frequency shift is determined precisely by the rf signal applied to the AOM. In this experiment these zeroth- and first-order beams are retroreflected (passed again) through the AOM resulting in overlapping zeroth- and second-order (shifted by twice the rf frequency) beams. Moreover, these beams can be selectively blocked leaving a resulting laser beam which is predominately of a single order. With this configuration in place, we direct only the unshifted zeroth-order retroreflected beam to the vacuum chamber where it excites the $F_1' = 1/2$, $F' = 0$ transition. The UV laser frequency is adjusted while monitoring the camera images until the pictorial resonance condition is achieved, signaling that the unshifted laser frequency is centered on that transition. The zeroth-order beam is then blocked and the second-order retroreflected beam is directed to the vacuum chamber. The AOM modulation frequency is then adjusted such that the second-order beam achieves the pictorial resonance condition for the $F_1' = 1/2$, $F' = 1$ transition. Double the AOM modulation frequency then gives the precise splitting between the two resonance lines. We confirm the laser has not drifted over the course of the measurement by again checking for pictorial resonance with the unshifted zeroth-order beam.

We measure five splittings in total with this method. We investigate four pairs of neighboring hyperfine levels by exciting either a pair of $R_0$ or a pair of $Q_1$ transitions. Specifically, we precisely measure the $\tilde{J} = 1$ hyperfine splittings between the $F' = F_1'+1/2$ and $F' = F_1'-1/2$ states for both $F_1' = 1/2$ and $F_1' = 3/2$; the $R_0$ transition gives information on the $P=-1$ manifold, and the $Q_1$ transition on the $P=+1$ manifold. In the language of Ref. [17], these correspond to the "$a$ separation" and "$c$ separation" of the $e$ and $f$ levels of $\tilde{J} = 1$, respectively. The fifth splitting is investigated by exciting the nearly degenerate $R_0$, $F_1' = 3/2$, $F' = 1$ and $Q_1$, $F_1' = 3/2$, $F' = 1$ transitions and adding to that measured frequency difference the rotational ground-state splitting (known to sub-MHz precision [15]). Our measured splittings agree with earlier results and are on average a factor of 4 more precise than those derived from the coefficients reported in Ref. [18] with uncertainties given by Ref. [24] (Table I). This recent analysis of Ref. [18] finds the HF splittings of $f$-parity/$e$-parity manifolds to be nearly identical, and so too do our AOM coincidence measurements.

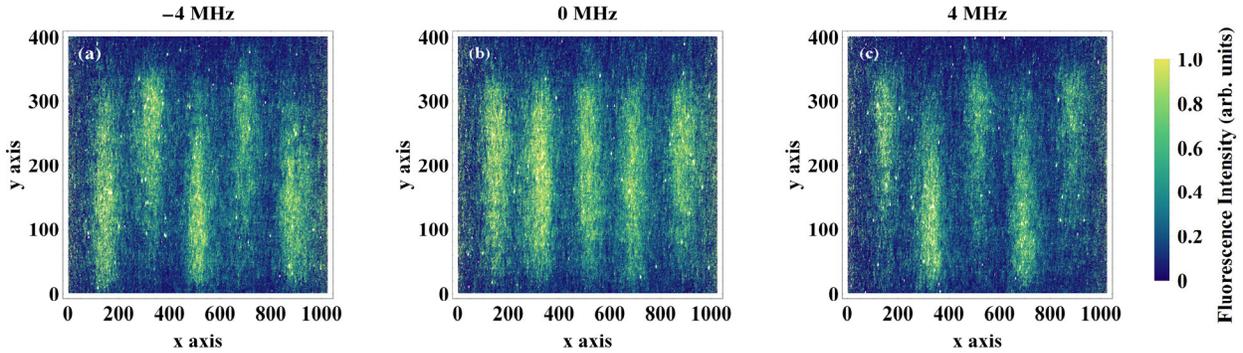

**FIG. 3.** Three fluorescence images of $Q_1$, $F_1' = 1/2$, $F' = 1$ transition excited by frequency-shifted laser light from the AOM. The AOM modulation frequency was varied between images in order to achieve the pictorial resonance condition. The molecular beam enters the multipass on the left side of the image moving in the $+x$ direction. With a total viewing area of ~18mm by ~7mm, each unit along the $x$- and $y$-axes corresponds to one camera pixel. Values above each frame specify how many MHz away from resonance each image was taken. Only when the fluorescence of all five passes occurs at the same location along the vertical axis, middle panel, is the laser frequency centered on the resonance. The natural linewidth of the transition is 1.6 MHz.

## B. Stark shift measurements

Our AOM coincident resonance measurements leave only two small unknown splittings labeled $\omega_1$ and $\omega_2$ for the $\tilde{J} = 1$ level of TlF (Fig. 4). These are referred to as Λ-doublet splittings for Hund's case (c) molecules and arise from perturbative coupling between electronic states [25]. In this section, we determine these splittings, as well as the excited $B$ state EDM $\mu_E$, via Stark shift spectroscopy and compare them to the results following from the data of Ref. [17], when correctly analyzed (Ref. [18]).

TlF is highly polarizable in the $^3\Pi_1$ state due to the nearly degenerate Λ-doublet levels of opposite parity. Electric fields of $E \sim 100$ V/cm can fully polarize molecules in this state. Spectra were obtained for four lines while varying an external electric field from 0-300 V/cm:

$$[J_g = 0^+\rangle \to [\tilde{J} = 1^-, F' = \tfrac{1}{2}, F' = 1\rangle, \quad (4)$$

$$[J_g = 0^+\rangle \to [\tilde{J} = 1^-, F' = \tfrac{3}{2}, F' = 1\rangle, \quad (5)$$

$$[J_g = 1^-\rangle \to [\tilde{J} = 1^+, F' = \tfrac{1}{2}, F' = 1\rangle, \quad (6)$$

$$[J_g = 1^-\rangle \to [\tilde{J} = 1^+, F' = \tfrac{3}{2}, F' = 1\rangle. \quad (7)$$

For these measurements, the laser polarization was modulated at 1 MHz between being perpendicular to and parallel to the electric field. For a given electric field, spectra were acquired by scanning the laser's frequency in 2-MHz increments and recording fluorescence signals (Fig. 5). The recorded peaks' line centers were determined by fitting Lorentzian functions to them. Other spectral line functions were considered, but empirically Lorentzians gave the best fits. The level structure of the $\tilde{J} = 1$, $F_1' = 1/2$, $F' = 1$ state in a strong electric field is shown in Fig. 6(a); the peak locations which vary with electric field, seen in Fig. 5, are a consequence of this structure. Figure 6(b) shows theoretical curves, obtained as described below, which give the relative frequency of these states as a function of electric field for the parameter values of this work. Figure 5 and Fig. 6 share the same color scheme for easy identification of peaks, and to make clear the naming conventions of these peaks.

Because the Doppler width and the Λ-doublet splitting are both of order 10 MHz, in the weak-field limit, the Λ-doublet splitting is not fully resolved. This leads to some ambiguity in fitting the spectra at low fields. Due to this ambiguity, the widths of the various peaks were equated in the fits. This was sufficient to constrain the spectra and resulted in convergent fits like those of Fig. 5. Furthermore, peaks separated by less than 1 full width half maximum have not been considered in this analysis due to the ambiguity associated with overlapping peaks.

An electric field in the $z$ direction, $\vec{E}_{Lab}$, mixes together the opposite parity states with the same total angular momentum projection $m_{F'}$. In the strong-field limit, $\Omega$, not $P$, is a good quantum number. To compensate for potential long-term drifts of our scanning Fabry-Pérot cavity, we take the difference frequency between the $m_{F'} \times \Omega = \pm 1$ and the $m_{F'} \times \Omega = 0$ line centers for each electric-field value. A spectrum for a single electric field takes typically only a few minutes to collect. For a given $F_1'$ value these differences, for both $R_0$ and $Q_1$, are plotted against electric field (Fig. 7). The measured Stark-shifted energies for $\tilde{J} = 1$ are then determined by the differences in fit $m_{F'} \times \Omega = \pm 1$ and the $m_{F'} \times \Omega = 0$ line centers and the observed field-free splittings listed in Table I, with $\mu_E$, $\omega_1$, and $\omega_2$ as free parameters. Note the diverging lines of Fig. 7 are not symmetric about zero because the $m_{F'} \times \Omega = 0$ line centers are shifted significantly at high fields due to mixing with the $F_1' = 1/2$, $F' = 0$ level as well (see Fig. 6).

| Label | Splitting | Ref. [18] (MHz) | This Work (MHz) |
|---|---|---|---|
| $a^+$ | $\tilde{J} = 1^+, F_1' = 1/2$ | 563(10) | 560(2) |
| $a^-$ | $\tilde{J} = 1^-, F_1' = 1/2$ | 563(10) | 562(2) |
| $c^+$ | $\tilde{J} = 1^+, F_1' = 3/2$ | 317(6) | 315(2) |
| $c^-$ | $\tilde{J} = 1^-, F_1' = 3/2$ | 316(6) | 315(2) |
| $z$ | $b^- + \omega_2 + c^+$ or $b^+ - \omega_1 + c^+$ | 501(7)+13 334.7 | 498(2)+13 334.7 |
| $b^+$ |  | 13 536(7) | 13 532(3) |
| $b^-$ |  | 13 502(7) | 13 500(3) |
| $\omega_1$ |  | 16.1(8) | 14.4(9) |
| $\omega_2$ |  | 17.7(13) | 17.4(11) |

Table I. Measured splittings using AOM coincidence resonance and dc Stark shift techniques. Uncertainties associated with Ref. [18] results have been obtained from Ref. [24].

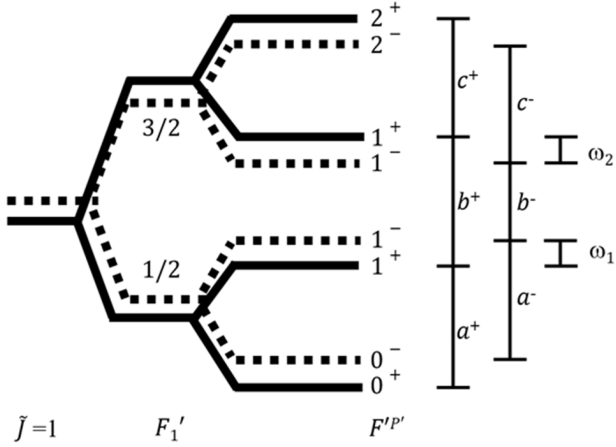

**FIG. 4.** Labeling scheme for the energy splittings in the $\tilde{J} = 1$ level of TlF. The solid (dashed) lines correspond to $f$-parity ($e$-parity) levels. The hyperfine structure in the ground state is completely unresolved and known to be <<1 MHz, so lines to all the depicted states of a given parity effectively originate from one common ground-state rotational level.

To theoretically describe our Stark-shifted spectra, we begin by calculating the field-free eigenenergies and eigenstates in the basis of Eq. 3 using the rotation and hyperfine interaction from Ref. [17], updated to include the small lambda-doubling contribution to the effective Tl nuclear spin-rotation parameter [26] as explicated in Ref. [18]. To fit our Stark shift measurements, we include the Stark Hamiltonian, $\mathcal{H}(t) = -T^1(\boldsymbol{E}) \cdot T^1(\boldsymbol{\mu}_e)$, from Eq. 6.318 of Ref. [25]. This has matrix elements between states with nuclear spin decoupled, given by Eq. 6.320 in Ref. [25]:

$$\langle J, \Omega, m_j | \boldsymbol{E} \cdot \boldsymbol{d} | J', \Omega', m_j' \rangle = -Ed(-1)^{2J - m_j - \Omega} \quad (8)$$
$$\times \begin{pmatrix} J & 1 & J' \\ \Omega & 0 & \Omega' \end{pmatrix} \begin{pmatrix} J & 1 & J' \\ -m_j & 0 & m_j' \end{pmatrix},$$

with $d$ equal to the excited $B$-state molecule-frame dipole moment, $\mu_E$, as a free parameter. For each applied electric field, we construct the $B$-state Hamiltonian in the $|J, F_1, F, P, m_F\rangle$ basis including $\tilde{J} = 1$-3 to account for the mixing of rotational levels by the applied field and hyperfine interaction.

The ground-state Stark shifts can be determined by diagonalizing the $X$-state Hamiltonian given in Ref. [1]. For the largest external fields used, the ground-state Stark shift is less than 10 MHz for the $J_g = 0, 1$ states. Because we take the difference between $m_{F'} \times \Omega = \pm 1$ and the $m_{F'} \times \Omega = 0$ line centers an average change in the ground-state energy has no effect on our analysis. Differential contributions from ground-state Stark splittings associated with state selection in the excitation process are mitigated by the modulation of the excitation polarization and are expected to be negligible at our current level of sensitivity.

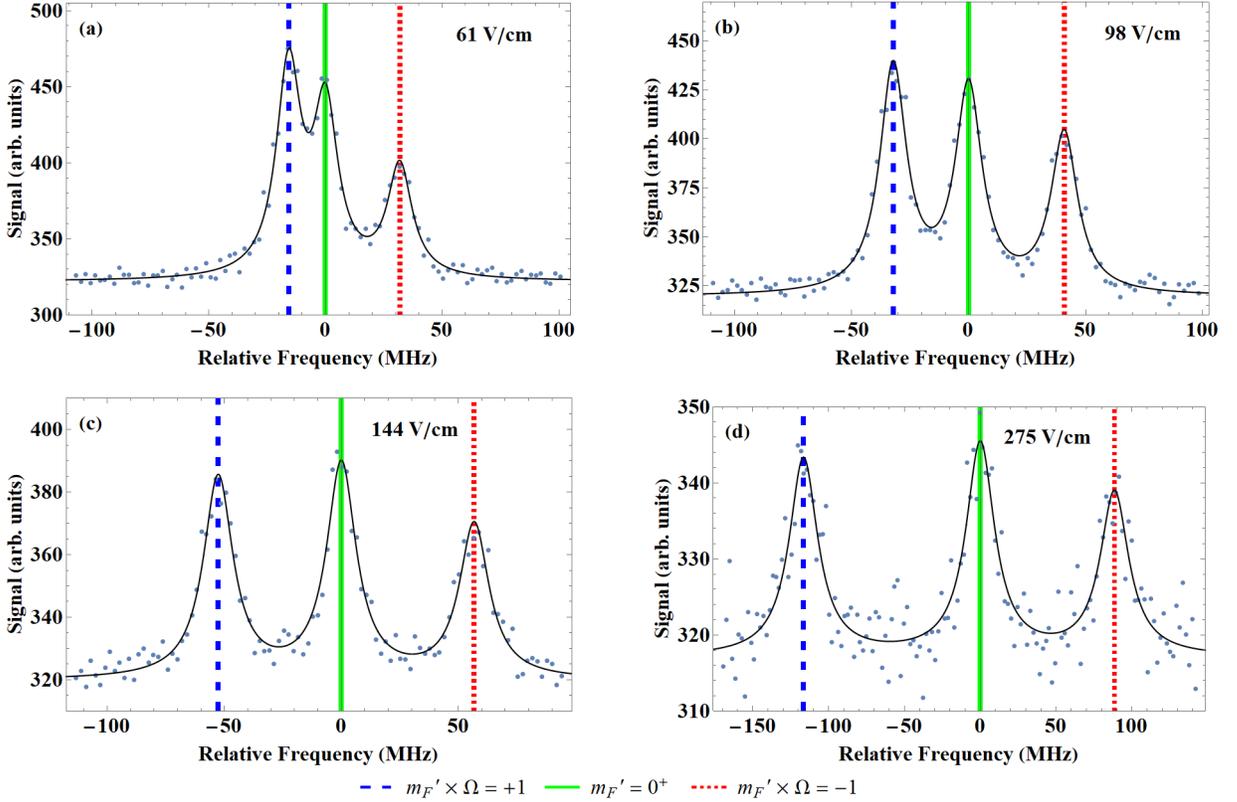

**FIG. 5.** Peaks of the $Q_1$, $F_1' = 1/2$, $F' = 1$ transition at various electric fields (denoted in the top right corner). The $m_F' = |1|$ states of opposite parity are mixed by the electric field. The frequency of the positive $m_F' \times \Omega$ (long dashes) and negative $m_F' \times \Omega$ (short dashes) states decrease and increase respectively with electric field. Because we take the difference frequency between the $m_F' \times \Omega = \pm 1$ and the $m_F' \times \Omega = 0$ line centers to generate the data points of Fig. 7(a), for a given electric field, the $m_F' = 0^+$ level (solid) corresponds to the zero ordinate value of Fig. 7(a). The upper $m_F' = 0^-$ level is not seen as it remains parity forbidden. A similar procedure using the relevant transitions generates the data points of the other transitions of Fig. 7.

The uncertainty in each $m_F' \times \Omega$ difference frequency is found by adding in quadrature the spectroscopic fit uncertainties of the relevant line centers. From finite-element simulations of our electric-field plates, we find the average electric field over the interaction region to be lower than predicted by an infinite capacitor model by 1.5%. The field is uniform to 94.7% over this region and the applied voltage is controlled with a precision of a few mV. We estimate the total instrumental uncertainty in the applied electric field to be 2%.

We determine the free parameters $\omega_1$, $\omega_2$, and $\mu_E$ by minimizing the root-mean-square difference between the measured and calculated energy splittings (Table II). Data and fitted curves are presented in Fig 7. We obtain $\omega_1$ = 14.4(8) MHz, $\omega_2$ = 17.4(10) MHz, and the $\mu_E$ = 2.28(3) D, where the numbers in parentheses is the 1 standard deviation confidence interval, $\delta$, for three fit parameters. The rms error was 1.6 MHz, with 141 degrees of freedom.

The systematic uncertainty has two major contributions, the uncertainty in the electric field and the uncertainty in the calibration of the Fabry-Pérot cavity used to determine the laser frequency. The systematic uncertainty associated with the electric field, $\delta_E$, is determined by assuming that a 2% instrumental uncertainty in the electric field corresponds to a 2% systematic uncertainty in $\mu_E$. We use this assumption, coupled with the covariance between $\mu_E$ and $\omega_{1,2}$ (as determined by the fits), to relate the electric-field instrumental uncertainty to a systematic uncertainty associated with the $\omega_{1,2}$ parameter measurements (Table II).

The nonlinear response of our Fabry-Pérot cavity has been investigated and modeled using repeated Fabry-Pérot cavity scans of our lasers and assuming the cavity has a fixed free spectral range. A conservative estimate of the systematic uncertainty associated with the cavity correction, $\delta_C$, has been determined and is recorded in Table II. Combining all of the contributing uncertainties in quadrature yields $\delta_{Total}$, and a final result: $\omega_1$ = 14.4(9) MHz, $\omega_2$ = 17.4(11) MHz, and $\mu_E$ = 2.28(7) D.

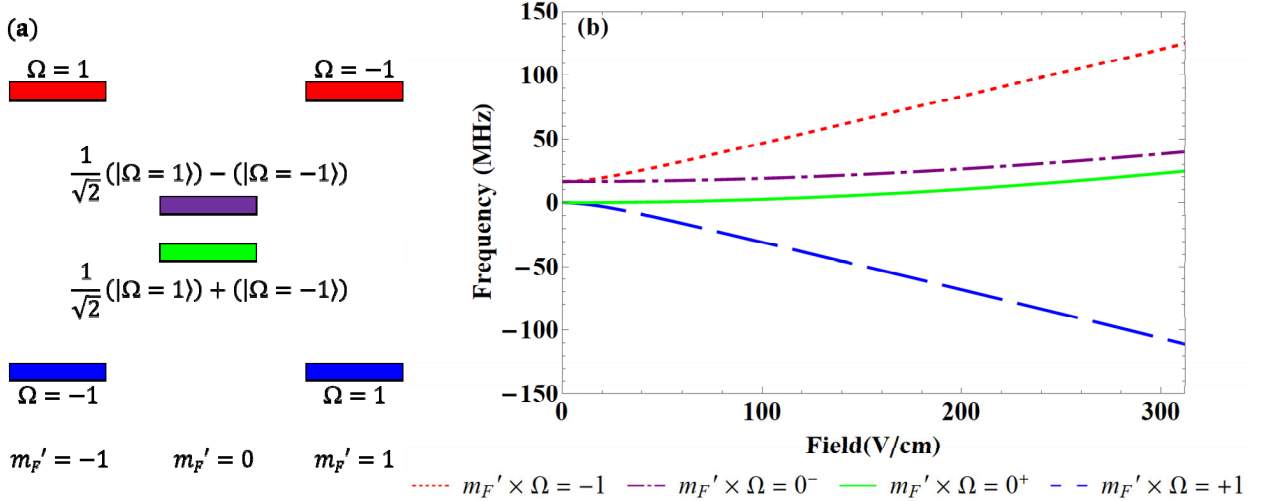

**FIG. 6.** (a) Level structure of the $\tilde{J}$ = 1, $F_1'$ = 1/2, $F'$ = 1 state in a strong electric field. The $m_F'$ = |1| states of opposite parity are mixed by the electric field. (b) Theoretical curves showing the relative frequencies of the $\tilde{J}$ = 1, $F_1'$ = 1/2, $F'$ = 1 states as a function of electric field assuming the splitting between nearest opposite-parity sublevels and dipole moment of this work. The frequency of the positive and negative $m_F' \times \Omega$ states decrease and increase respectively, with electric field (see Fig. 5). Note also that the $m_F'$ = $0^+$ and $m_F'$ = $0^-$ states' frequencies increase slightly with increasing electric field due to mixing with the $F_1'$ = 1/2, $F'$ = 0 level.

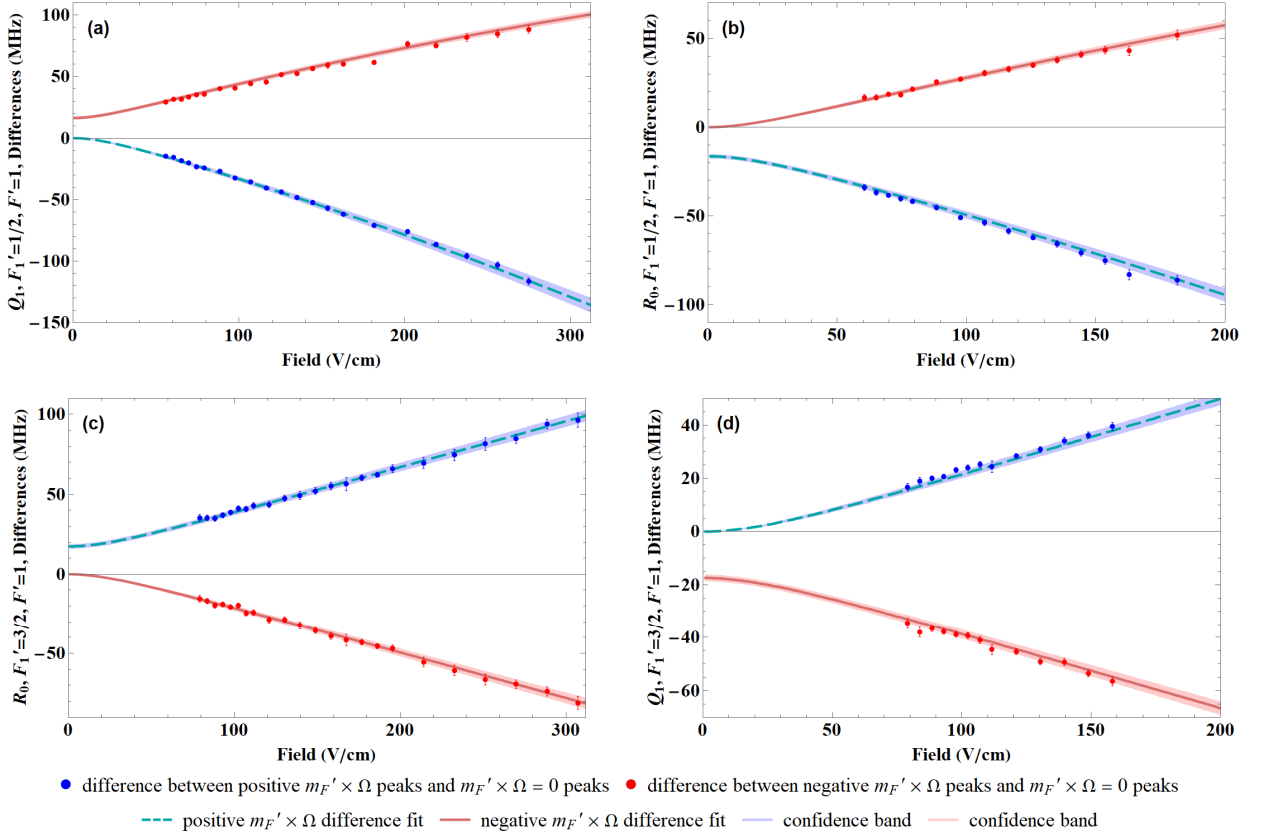

**FIG. 7.** Experimentally measured differences between the positive $m_F' \times \Omega$ peaks or negative $m_F' \times \Omega$ peaks and $m_F' \times \Omega = 0$ peaks for different transitions. Curves are error bands determined by varying the $\Omega$-splittings and EDM values within their respective 1-sigma confidence intervals. Error bars on individual points are indicated.

| Parameter | Units | Value | $\delta$ | $\delta_E$ | $\delta_C$ | $\delta_{Total}$ |
|---|---|---|---|---|---|---|
| $\omega_1$ | MHz | 14.4 | 0.8 | 0.1 | 0.4 | 0.9 |
| $\omega_2$ | MHz | 17.4 | 1.0 | 0.1 | 0.4 | 1.1 |
| $\mu_E$ | D | 2.28 | 0.03 | 0.05 | 0.04 | 0.07 |

**Table II.** Stark measurement parameters with assigned statistical and systematic uncertainties (see text).

## V. CONCLUSIONS

We have observed the spatially resolved images of molecular fluorescence produced from the multiple passage of a narrow-band laser beam through an orthogonal cryogenic molecular beam of TlF. From these images, and using an acoustic optic modulator for frequency calibration, we have determined several TlF $B$-state hyperfine splittings with uncertainties that are on average a factor of 4 smaller than those previously reported in Ref. [18]. Combining these refined splittings with Stark shift measurements we have determined the molecule-frame dipole moment $\mu_E$, as well as the $\Lambda$-doublet splittings, $\omega_1$ and $\omega_2$, between the opposite parity $\tilde{J} = 1$, $F_1' = 1/2, 3/2$ levels, respectively. Our $\Lambda$-doublet results are in reasonably good agreement with the values reported in Ref. [18] and we find the same energy ordering. We are unaware of any previous measurements of $\mu_E$ for the TlF B state. Furthermore, these results by direct measurement agree with the energy ordering of $f$-parity and $e$-parity states in $\tilde{J} = 1$ determined by the analysis of Ref. [18].

The hyperfine mixing of excited-state rotational levels is an important consideration for optical cycling and laser-cooling applications in TlF as it can lead to excess rotational branching compared to a nuclear-spin zero system [5,17]. Because the magnetic hyperfine interaction due to the Tl nuclear spin in the TlF $B$ state [$h_1$(Tl) = 28 789 MHz] is similar in magnitude to the rotational interaction ($B$ = 6688 MHz), states are strongly mixed when they have the same values of $F_1'$ and $P$, and $\tilde{J}$ differing by 1. This mixing is notably absent in in $\tilde{J} = 1$, $F_1' = 1/2$ as there is no other $F_1' = 1/2$ level. It is critical to include the Tl magnetic hyperfine interaction to accurately predict the Stark-shifted spectra, as it is stronger than the Stark interaction even for the largest fields we applied. The fact that the combined Stark spectra for the $\tilde{J} = 1$, $F_1' = 1/2$, $F' = 1$ level (no Tl magnetic hyperfine mixing) and in $\tilde{J} = 1$ $F_1' = 3/2$, $F' = 1$ (strong Tl magnetic hyperfine mixing with $\tilde{J} = 2$) can be described by the same $\mu_E$ with 3% fractional statistical uncertainty strongly validates the assigned hyperfine parameters and their predicted state mixing [17,18].

Overall, we have achieved a more complete understanding of the hyperfine interactions which significantly mix neighboring rotational levels and thus allow additional rotational branching from the $B^3\Pi_1$ state. Parity mixing from stray or residual electric fields induces further rotational branching which is only strictly parity forbidden in zero field. Our experimentally determined values of $\omega_1$, $\omega_2$, and $\mu_E$ can set quantitative limits on the magnitude of electric fields permissible in order to scatter a desired number of photons in optical cycling applications. The TlF molecule remains an interesting candidate for symmetry-violation measurements, and potentially laser cooling and trapping.

## ACKNOWLEDGMENTS

This work was supported by National Science Foundation Grants No. PHY 1519265, No. PHY 1806297, and by the Heising-Simons Foundation. E.B.N. acknowledges support from National Institute of Standards and Technology and the National Research Council Postdoctoral Research Associateship Program. The authors thank O. Grasdijk and J. Kastelic for producing TlF targets, O. Timgren for useful discussion and assistance with target fabrication, and B. J. Crepeau, J. M. Kubasek, and S. K. Peck for technical support.This work was supported by National Science Foundation Grants No. PHY 1519265, No. PHY 1806297, and by the Heising-Simons Foundation. E.B.N. acknowledges support from National Institute of Standards and Technology and the National Research Council Postdoctoral Research Associateship Program. The authors thank O. Grasdijk and J. Kastelic for producing TlF targets, O. Timgren for useful discussion and assistance with target fabrication, and B. J. Crepeau, J. M. Kubasek, and S. K. Peck for technical support.

# REFERENCES

[1] E. A. Hinds and P. G. H. Sandars, Phys. Rev. A **21**, 480 (1980).
[2] D. A. Wilkening, N. F. Ramsey, and D. J. Larson, Phys. Rev. A **29**, 425 (1984).
[3] D. Cho, K. Sangster, and E. A. Hinds, Phys. Rev. A **44**, 2783 (1991).
[4] L. I. Schiff, Phys. Rev. **132**, 2194 (1963).
[5] L. R. Hunter, S. K. Peck, A. S. Greenspon, S. S. Alam, and D. DeMille, Phys. Rev. A **85**, 012511 (2012).
[6] E. S. Shuman, J. F. Barry, D. R. Glenn, and D. DeMille, Phys. Rev. Lett. **103**, 223001 (2009).
[7] E. S. Shuman, J. F. Barry, and D. DeMille, Nature (London) **467**, 820 (2010).
[8] J. F. Barry, D. J. McCarron, E. N. Norrgard, M. H. Steinecker, and D. DeMille, Nature (London) **512**, 286 (2014).
[9] P. Aggarwal *et al.* (The NL-eEDM Collaboration), Eur. Phys. J. D **72**, 197 (2018).
[10] T. A. Isaev, S. Hoekstra, and R. Berger, Phys. Rev. A **82**, 052521 (2010).
[11] I. Smallman, F. Wang, T. Steimle, M. Tarbutt, and E. A. Hinds, J. Mol. Spectrosc. **300**, 3 (2014).
[12] J. Lim, J. R. Almond, M. A. Trigatzis, J. A. Devlin, N. J. Fitch, B. E. Sauer, M. R. Tarbutt, and E. A. Hinds, Phys. Rev. Lett. **120**, 123201 (2018).
[13] M. Denis, Pi A. B. Haase, R. G. E. Timmermans, E. Eliav, N. R. Hutzler, and A. Borschevsky, Phys. Rev. A **99**, 042512 (2019).
[14] E. B. Norrgard, D. S. Barker, S. Eckel, J. A. Fedchak, N. N. Klimov, and J. Scherschligt, Comm. Phys **2**, 77 (2019).
[15] J. Hoeft, F. J. Lovas, E. Tiemann, and T. Törring, Z. Naturforsch. A **25**, 1029 (1970).
[16] U. Wolf and E. Tiemann, Chem. Phys. Lett. **133**, 116 (1987).
[17] E. B. Norrgard, E. R. Edwards, D. J. McCarron, M. H. Steinecker, D. DeMille, S. S. Alam, S. K. Peck, N. S. Wadia, and L. R. Hunter, Phys. Rev A **95**, 062506 (2017).
[18] G. Meijer, and B. G. Sartakov, Phys. Rev. A **101**, 042506 (2020).
[19] J. F. Barry, E. S. Shuman, and D. DeMille, Phys. Chem. Chem. Phys. **13**, 18936 (2011).
[20] Any mention of commercial products within this work is for information only; it does not imply recommendation or endorsement by NIST.
[21] A. L. Vitushkin and L. F. Vitushkin, Appl. Opt. **37**, 162 (1998).
[22] D. J. Berkeland and M. G. Boshier, Phys. Rev. A **65**, 033413 (2002).
[23] G. Herzberg, *Molecular Spectra and Molecular Structure: I. Spectra of Diatomic Molecules*, 2nd ed. (Van Nostrand, New York, 1950).
[24] G. Meijer and B. G. Sartakov (private communication).
[25] J. M. Brown and A. Carrington, Rotational Spectroscopy of Diatomic Molecules (Cambridge University Press, Cambridge, UK, 2003).
[26] J. M. Brown, M. Kaise, C. M. L. Kerr, and D. J. Milton, Mol. Phys. **36**, 2, 553-582 (1978).